\documentstyle[twocolumn,epsf,floats,aps,prb]{revtex}
\begin{document}
\draft

\twocolumn[\hsize\textwidth\columnwidth\hsize\csname @twocolumnfalse\endcsname

\title{Correlated hopping in the Falicov-Kimball model:
       A large-dimensions study}
\smallskip
\author{Avraham Schiller}
\address{Racah Institute of Physics, The Hebrew University,
      Jerusalem 91904, Israel}
\date{\today}
\maketitle
\smallskip

\begin{abstract}
The Falicov-Kimball model with a correlated-hopping interaction is
solved using an extended dynamical mean-field theory that becomes
exact in the limit of large dimensions. The effect of correlated
hopping is to introduce nonlocal self-energy components that retain
full dynamics as $D \to \infty$, thus introducing an explicit
$k$-dependence to the single-particle self-energy. An explicit
solution for the homogeneous phase at $D = 2$
reveals significant nonlocal dynamical contributions in the physically
relevant regime of a moderately large correlated-hopping amplitude,
indicating that important nonlocal correlations are omitted in
Hubbard-like models upon neglecting the correlated-hopping interaction.
\end{abstract}

\smallskip

]
\narrowtext

The development of the limit of infinite dimensions,~\cite{MV89} or the
Dynamical Mean-Field Theory (DMFT), has opened exciting new possibilities
in the study of correlated electron systems.~\cite{GKKR96} One notable
example is the Mott-Hubbard metal-insulator transition realized in
V$_{2-y}$O$_3$ and Ca$_{1-x}$Sr$_x$VO$_3$, where the main features of the
optical conductivity and the photoemission data are successfully
accounted for within the DMFT treatment of the single-band Hubbard
model.~\cite{V2O3-Thomas,CaSrVO3-I} This approach, however, fails to
explain the moderate mass enhancement and the apparent $k$-dependence
of the single-particle self-energy close to the transition in these
materials,~\cite{V2O3-Kim,CaSrVO3-II} indicating that important
nonlocal interactions are either omitted in the Hubbard model, or
mistreated in the DMFT.

One nonlocal interaction term clearly absent in the Hubbard model is
that of correlated hopping (CH), in which hopping between neighboring
lattice sites depends on the occupancy of the opposite spin orientation.
Such an interaction term is an integral part of the
Coulomb repulsion between electrons, and with estimates ranging from about
$0.5{\rm eV}$ in transition metals,~\cite{Hubbard63} to $0.8{\rm eV}$
in the cuperates,~\cite{AGP93} and up to $3.3{\rm eV}$ in
benzene,~\cite{benzene} it is certainly comparable to, if not larger than,
the corresponding amplitude for single-particle hopping in real systems.
Despite this fact and in spite of Hirsch's suggestion of
a new mechanism for
superconductivity,~\cite{Hirsch89} there is still no qualitative
understanding of the effects that a CH term might have.
While some rigorous statements can be made, these are restricted to
certain models in one spatial
dimension,~\cite{Aligia_etal,Karn94_BKSZ95,SSZ97,dBKS95} and to a few
very special cases in higher dimensions.~\cite{dBKS95,SV93,Ovch93}
There is thus an obvious need for a careful investigation of CH within
a reliable nonperturbative approach, applicable to the entire range of
interactions from weak to strong coupling.

In this paper, we generalize the limit of infinite dimensions, $D \to
\infty$, as to include the CH interaction. Focusing on
the Falicov-Kimball model~\cite{FK69} --- a simplified Hubbard model
with one spin species tied down --- the effect of CH is to introduce
{\em nonlocal} self-energy components that
retain full dynamics as $D \to \infty$. This marks departure
from the local approximation. Similar to existing formulations
of the infinite-dimensional limit, though, both the local and nonlocal
self-energy components are extracted from an effective single-site
action, which can still be treated analytically for the Falicov-Kimball
model. Solving the resulting DMFT for the homogeneous phase
at $D = 2$ and in the physically
relevant regime of a moderately large CH amplitude, the dynamical
contribution of the nearest-neighbor self-energy is found to
be comparable to that of the local self-energy, indicating that
important nonlocal correlations are omitted in Hubbard-like models
upon neglecting the CH interaction. This not only suggests that CH may
be essential for the complete understanding of the metal-insulator
transition in V$_{2-y}$O$_3$ and Ca$_{1-x}$Sr$_x$VO$_3$, but
further calls into question its omission in any Hubbard-like
modeling of real systems.

We begin our discussion with the Hamiltonian
\begin{eqnarray}
{\cal H} &=& \frac{t_1}{\sqrt{2D}}\sum_{<i,j>}
         \left\{ d^{\dagger}_i d_j + d^{\dagger}_j d_i \right\} +
         U\sum_i d^{\dagger}_i d_i f^{\dagger}_i f_i
\nonumber\\
     &+& \frac{t_2}{\sqrt{2D}}\sum_{<i,j>}
         \left\{ d^{\dagger}_i d_j + d^{\dagger}_j d_i \right\}
         \left( f^{\dagger}_i f_i + f^{\dagger}_j f_j \right)
\nonumber\\
     &+& ( E_f - \mu )\sum_i f^{\dagger}_i f_i 
     - \mu \sum_i d^{\dagger}_i d_i ,
\label{H}
\end{eqnarray}
in which itinerant $d$ fermions interact via an on-site Coulomb repulsion
$U$ with localized $f$ fermions on a $D$-dimensional hypercubic lattice.
Motion of the $d$ fermions has two components: a single-particle,
nearest-neighbor hopping term $t_1$, and a correlated-hopping term $t_2$,
in which the hopping amplitude for the $d$ fermions depends on the
occupancy of the
corresponding $f$ fermions. The two fermion species share a joint
chemical potential, $\mu$.

For $t_2 = 0$, Eq.~(\ref{H}) reduces to the Falicov-Kimball
model,~\cite{FK69} which is perhaps the simplest many-body
Hamiltonian to exhibit both a Mott-Hubbard metal-insulator transition
and long-range ordered phases. A nonzero $t_2$ modifies
the single-particle dynamics of the itinerant fermions in two ways.
First, there is a straightforward Hartree renormalization of the
static single-particle hopping amplitude according to $t_1 \to t
= t_1 + 2 n_f t_2$, where $n_f$ is the $f$ occupation number (assuming
a homogeneous phase). In addition, the $d$ and $f$ fermions acquire
new self-energy terms with full dynamics, which survive also in the
absence of an on-site $U$. To investigate the
effect of CH we employ the limit of large dimensions,~\cite{MV89}
which has proven useful in studying the Falicov-Kimball
model.~\cite{Brandt,Freericks} To this end, both the
single-particle and CH terms in Eq.~(\ref{H}) have been properly
rescaled with $D$.

For conventional Hubbard-like models, the self-energy is purely
local at $D = \infty$, rendering the local approximation exact.
This paradigm breaks down as soon as CH is considered,
as exemplified in Fig.~1 for the Hamiltonian of Eq.~(\ref{H}).
In Fig.~1, sites $j$ and $k$ are distinct nearest neighbors
of site $i$. Thus, after summation over all equivalent $k$, the
diagram of Fig.~1(a) gives a contribution of order $1/\sqrt{D}$
to the nearest-neighbor $d$ self-energy $\Sigma_{ij}(z)$, while the
diagram of Fig.~1(b) (in which both $j$ and $k$ are fixed) gives a
contribution of order $1/D$ to the next-nearest-neighbor self-energy
$\Sigma_{kj}(z)$. Since both orders in $D$ coincide with the necessary
large-$D$ rescaling of the corresponding nearest- and next-nearest-neighbor
hopping amplitudes,~\cite{Mueller-Hartmann} both self-energies contribute
in the limit $D \to \infty$. As shown below, the nearest-neighbor
and next-nearest-neighbor self-energies are in fact the only nonlocal
self-energy components relevant at $D = \infty$.

\begin{figure}
\centerline{
\vbox{\epsfxsize=85mm \epsfbox {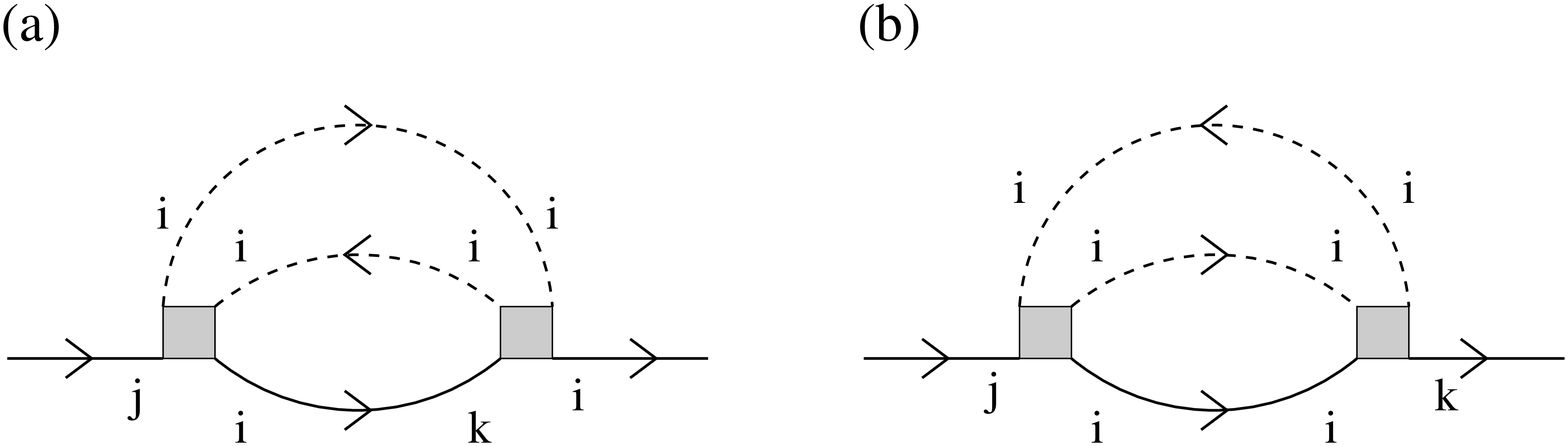}}
}\vspace{5pt}
\caption{Representative $d$ self-energy diagrams. Here, squares
indicate a correlated-hopping vertex ($t_2$); full (dashed)
lines represent bare $d$ ($f$) propagators; and $j$ and $k$ are
distinct nearest neighbors of site $i$. In Fig.~(a), there is
summation over all equivalent sites $k$. The diagram of Fig.~(a)
gives a contribution of order $D^{-1/2}$ to $\Sigma_{ij}$, while
the diagram of Fig.~(b) gives a contribution of order $D^{-1}$
to $\Sigma_{kj}$. Both terms are relevant in the limit
$D \to \infty$.}
\end{figure}

A systematic formulation of the $D \to \infty$ limit for the
Hamiltonian of Eq.~(\ref{H}) employs the Luttinger-Ward
functional $\Phi$ --- the sum of all vacuum-to-vacuum skeleton
diagrams. Consider a generic $\Phi$ diagram. Labeling each
vertex according to the site index of its incoming and outgoing
$f$ lines, we attach the $1/\sqrt{D}$ factor associated with each
CH vertex to its incoming or outgoing nearest-neighbor
$d$ line. With this convention we notice that, even after summation
over all intermediate sites, any path connecting two vertices at
sites $i \neq j$ gives a contribution of order
$D^{-\parallel i - j \parallel / 2}$ or higher to the diagram at
hand ($\parallel\!\!i - j\!\!\parallel$ is the minimal number
of nearest-neighbor steps leading from site $i$ to site $j$). Given
that (i) any two vertices within a $\Phi$ diagram can be connected by
four independent paths having no lines in common, and (ii) the total
number of equivalent sites $j$ scales as $D^{\parallel i - j \parallel}$,
the overall contribution of such a diagram (after summation over all
equivalent $j$'s) is ${\cal O} \left( D^{-\parallel i - j
\parallel} \right)$. Thus, to order $D^{-n}$, only
skeleton diagrams with a maximal inter-vertex distance
$\parallel\!\!i - j\!\!\parallel_{max} \le n$ need to be considered.
Specifically, only purely local vertices are left at $D = \infty$.

While the above classification appears identical to that for
conventional Hubbard-like models, it should be emphasized that purely
local vertices in our convention do not imply purely local diagrams.
To the contrary, a CH vertex at site $i$ couples the operators $d_i$
and $f_i$ to the nearest-neighbor shell operator
\begin{equation}
\psi_i \equiv \frac{1}{\sqrt{2D}}
       \sum_{<j,i>} d_{j} ,
\end{equation}
in which $j$ is summed over the $2D$ nearest neighbors
of site $i$. Thus, a CH vertex at site $i$ necessarily
introduces at least one propagator that is not confined to site $i$.

Despite the above nonlocality, the $D \to \infty$ limit remains
tractable since $\Phi$ decouples in this limit into
\begin{equation}
\Phi = \sum_i \Phi_{loc}[G^{dd}_{ii}, G^{d \psi}_{ii},
       G^{\psi d}_{ii}, G^{\psi \psi}_{ii}, G^{ff}_{ii}] ,
\label{Phi}
\end{equation}
where $\Phi_{loc}$ is the functional generated by the action
\begin{eqnarray}
S_{eff} = &-&\!\!\int_0^{\beta} \!\!d\tau \int_0^{\beta} \!\!d\tau'\!\!
       \sum_{\alpha, \beta = d, \psi}\!\! \alpha^{\dagger}(\tau)
       \left [ {\cal G}^{-1} (\tau-\tau') \right ]_{\alpha\beta}
       \beta(\tau')
\nonumber\\
   &-& \!\!\int_0^{\beta} \!\!d\tau \int_0^{\beta}
       \!\!d\tau' f^{\dagger}(\tau) {\cal G}_f^{-1}
       (\tau-\tau') f(\tau')
\nonumber\\
   &+& t_2 \int_{0}^{\beta}\!\! d\tau
       \left \{ d^{\dagger}(\tau) \psi(\tau) + 
       \psi^{\dagger}(\tau) d(\tau) \right \} 
       f^{\dagger}(\tau)f(\tau) 
\nonumber\\
   &+& U\int_{0}^{\beta}\!\! d\tau \; d^{\dagger}(\tau)
       d(\tau) f^{\dagger}(\tau)f(\tau) .
\label{effective_S}
\end{eqnarray}
Here ${\cal G}(\tau - \tau')$ is a $2 \times 2$ matrix propagator,
chosen such that, when dressed with the effective-action self-energy,
it coincides with the corresponding lattice propagator:
\begin{equation}
{\cal G}^{-1}(i\omega_n) = G^{-1}(i\omega_n) +
      \Sigma^{S}(i\omega_n)
\label{g_definition}
\end{equation}
with
\begin{equation}
G(i\omega_n) = \left [
     \begin{array}{cc}
           G^{dd}_{ii}(i\omega_n) , & G^{d \psi}_{ii}(i\omega_n) \\ \\
           G^{\psi d}_{ii}(i\omega_n) , & G^{\psi \psi}_{ii}(i\omega_n)
     \end{array} \right ]
\label{G_definition}
\end{equation}
[$\omega_{n} = \pi (2n + 1) T$ are the Matsubara frequencies]. Here,
in using the $2 \times 2$ matrix notation of
Eqs.~(\ref{g_definition})--(\ref{G_definition}), we have identified
the indices $\alpha = d, \psi$ with $\alpha = 1, 2$, respectively.
Contrary to ${\cal G}$, which requires knowledge of the fully dressed
lattice $d$ propagator, the bare $f$ propagator in Eq.~(\ref{effective_S})
is simply ${\cal G}_f^{-1}(i\omega_n) = i\omega_n + \mu - E_f$, which
stems from the localized nature of the $f$ fermions.

From Eq.~(\ref{Phi}) it is apparent that $\Phi$ at $D = \infty$
is a functional of the local $d$ and $f$ propagators, as well as
the nearest-neighbor and next-nearest-neighbor $d$ Green functions
which enter via $G^{d\psi}_{ii}, G^{\psi d}_{ii}$, and $G^{\psi\psi}_{ii}$.
Thus, in addition to the local $d$ and $f$ self-energies, one must
also consider the nearest- and next-nearest-neighbor $d$ self-energies.
From a functional derivative of Eq.~(\ref{Phi}) one has
\begin{mathletters}
\label{self_energies}
\begin{eqnarray}
&&\Sigma_{ii}(i\omega_n) = \Sigma^{S}_{dd}(i\omega_n) 
      + \Sigma^{S}_{\psi \psi}(i\omega_n) ,\\
&&\Sigma_{<i,j>}(i\omega_n)  =
      \frac{1}{\sqrt{2D}} \left [ \Sigma^{S}_{d\psi}(i\omega_n) 
      + \Sigma^{S}_{\psi d}(i\omega_n) \right] ,\\
&&\Sigma_{<\!\!<\!i,j\!>\!\!>}(i\omega_n) =
      \frac{1}{2D} N_{<\!\!<\!i,j\!>\!\!>}
      \Sigma^{S}_{\psi\psi}(i\omega_n) ,
\label{nnn_self_energy}
\end{eqnarray}
\end{mathletters}

\noindent where $<\!\!<\!\!i,j\!\!>\!\!>$ denotes distinct lattice
sites with a common nearest neighbor. In Eq.~(\ref{nnn_self_energy}),
$N_{<\!\!<\!i,j\!>\!\!>}$ is equal to one for $i$ and $j$ on
the same axis; otherwise it is equal to two. Accordingly,
the $\vec{k}$-dependent $d$ self-energy reads
\begin{equation}
\Sigma_{\vec{k}}(z) =
      \Sigma^{S}_{dd}(z) + \epsilon_{\vec{k}}
      \left[ \Sigma^{S}_{d\psi}(z) + \Sigma^{S}_{\psi d}(z) \right]
      + \epsilon_{\vec{k}}^2 \Sigma^{S}_{\psi\psi}(z) ,
\label{sigma_k}
\end{equation}
where $\epsilon_{\vec{k}} = \sqrt{2/D}\sum_{l = 1}^{D} \cos(k_l)$.
Notice that, contrary to conventional formulations of the limit
of large dimensions, $\Sigma_{\vec{k}}$ has an explicit ${\vec{k}}$
dependence which enters solely through the dimensionless energy
${\epsilon_{\vec{k}}}$:
$\Sigma_{\vec{k}}(z) \equiv \Sigma_{\epsilon_{\vec{k}}}(z)$.

A self-consistency loop can finally be closed by expressing the
lattice Green function of Eq.~(\ref{G_definition}) in terms of
$\Sigma_{\vec{k}}$. For the spatially homogeneous phase
this gives
\begin{equation}
G(z) = \int_{-\infty}^{\infty} d\epsilon \frac{\rho(\epsilon)}
       {z + \mu - t_1 \epsilon - \Sigma_{\epsilon}(z)} \left [
       \begin{array}{cc}
           1 , & \epsilon \\ \\
           \epsilon , & \epsilon^2
     \end{array} \right ] ,
\label{G_lattice}
\end{equation}
where $\rho(\epsilon)$ is the density of states for the energy
$\epsilon_{\vec{k}}$, i.e., a Gaussian for $D = \infty$.
Equation~(\ref{G_lattice}) can further be written in closed form
in terms of the bare ($U, t_2 = 0$) tight-binding $d$ propagator.
We also note that Eqs.~(\ref{sigma_k})--(\ref{G_lattice}) were
derived under the explicit assumption of a hypercubic lattice
with only nearest-neighbor single-particle hopping. Different 
self-consistency equations will generally apply to other lattices
and other tight-binding models.

Equations~(\ref{effective_S})--(\ref{G_lattice}) are an exact
formulation of the limit $D \to \infty$ for the Hamiltonian
of Eq.~(\ref{H}). In particular,
for $t_2 = 0$, when only $\Sigma^S_{dd}(i\omega_n)$
is nonzero, they properly reduce to the corresponding equations for
the Falicov-Kimball model at $D = \infty$.~\cite{Brandt} In the spirit
of the DMFT, the same set of equations can also be applied to any finite
dimension $D$, provided the actual $D$-dimensional density of states
$\rho(\epsilon)$ is used in Eq.~(\ref{G_lattice}). Below we
present results for $D = 2$.
Although designed at present to describe the homogeneous phase, this
formulation is easily extended to phases with long-range order,
as well as to other lattice models with CH.
For example, in the Hubbard model one has to introduce separate
$\psi_{i\sigma}$ fields for each spin orientation,
but the basic formulation remains the same.

A crucial advantage of the Hamiltonian of Eq.~(\ref{H}) over its Hubbard
counterpart is that the action of Eq.~(\ref{effective_S}) can be treated
analytically. Integrating out the $f$ particles, the
lattice Green function of Eq.~(\ref{G_definition}) is expressed as
\begin{equation}
G(i\omega_{n}) = (1\!-\!n_f) {\cal G}(i\omega_{n}) + n_f\!
     \left [ {\cal G}^{-1}(i\omega_{n}) - \left (\!
\begin{array}{cc}
U & t_2 \\
t_2 & 0
\end{array}
\!\right )\right ]^{-1}\! ,
\label{solution_of_G}
\end{equation}
where $n_f$ is the $f$ occupation number:
$n_f = 1/(e^{\beta \epsilon} + 1)$,
\begin{equation}
\epsilon = E_f - \mu + U/2 - \frac{1}{\beta}
        \sum_{n=-\infty}^{\infty}\!\! \ln {\rm det}\!\!
          \left [ 1\! - {\cal G}(i\omega_{n}) \left (\!
              \begin{array}{cc}
              U & t_2 \\
              t_2 & 0
              \end{array}
          \!\right )\right ]\! .
\label{ln_Z_1}
\end{equation}
Once Eqs.~(\ref{g_definition}), (\ref{G_lattice})--(\ref{ln_Z_1}) have
been iterated and the $f$ occupation number determined, the same
equations can be used to calculate $G(z)$ for an arbitrary complex $z$.

\begin{figure}
\centerline{
\vbox{\epsfxsize=80mm \epsfbox {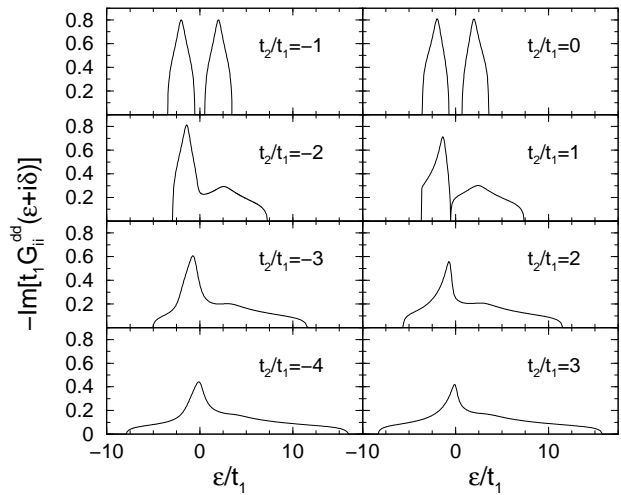}}
}\vspace{2pt}
\caption{The $d$ spectral function as a function of $t_2/t_1$,
for $n_f + n_d = 1$, $U/t_1 = 4$, $T/t_1 = 1$, and $D = 2$. For
both $t_2/t_1 = 0$ and $t_2/t_1 = -1$, the half-filled model is
particle-hole symmetric, with a Mott gap for $U/t_1 = 4$. This
gap gradually closes upon increasing $t_2/t_1 >0$ or decreasing
$t_2/t_1 < -1$, until the lower and upper Hubbard bands are merged
into one broad band. Note the general resemblance between curves with
values of $t_2/t_1$ that are symmetric about $t_2/t_1 = -1/2$.}
\end{figure}

Restricting attention to $D = 2$ and $E_f = 0$, we focus hereafter on
the homogeneous phase of the half-filled case: $n_f + n_d = 1$, where
$n_d$ is the $d$ occupation number. Setting $U/t_1 = 4$ and $T/t_1 = 1$,
Fig.~2 shows the evolution of the $d$ spectral function as a function
of $t_2/t_1$. For the Falicov-Kimball model, $t_2 = 0$, the $d$ spectrum
has a Mott gap. This gap persists throughout the range
$-1.4 \alt t_2/t_1 \alt 1$,
although the chemical potential does not always fall inside the gap.
For either $t_2/t_1 \agt 1$ or $t_2/t_1 \alt -1.4$, the lower and upper
Hubbard bands merge into one broad band, reflecting the increase in
kinetic energy of the itinerant fermions. Due to the lack of particle-hole
symmetry, the $d$ spectrum is generally asymmetric. Indeed, only
for $t_2 = 0$ and $t_2/t_1 = -1$ is the half-filled model particle-hole
symmetric, in which case the resulting spectrum is temperature
independent.~\cite{PH-comment}

\begin{figure}
\centerline{
\vbox{\epsfxsize=65mm \epsfbox {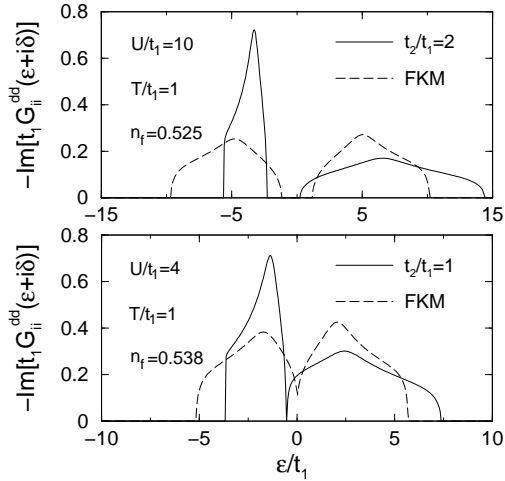}}
}\vspace{15pt}
\caption{The $d$ spectrum for the Hamiltonian of Eq.~(\protect\ref{H})
at $D = 2$ versus the one obtained within the DMFT for the Falicov-Kimball
model with the renormalized single-particle hopping,
$t_{FKM} = t_1 + 2 t_2 n_f$. Here $n_f$ and $n_d = 1 - n_f$ are kept the
same in both models, as are $U/t_1$ and $T/t_1$.}
\end{figure}

As shown in Fig.~3, one cannot simply absorb $t_2$ into an effective
Hartree renormalization of the single-particle hopping within the simpler
Falicov-Kimball model. Specifically, tuning $\mu$ and $E_f$ in the
latter model as to maintain the same $n_f$ and $n_d$,
there are substantial deviations between the $d$ spectral function
for the Hamiltonian of Eq.~(\ref{H}) and that obtained within the
DMFT for the Falicov-Kimball model with the renormalized hopping,
$t_{FKM} = t_1 + 2 t_2 n_f$. The $d$ spectrum for a nonzero $t_2$ is
notably more asymmetric, with considerable broadening (narrowing)
of the upper (lower) Hubbard band.

\begin{figure}
\centerline{
\vbox{\epsfxsize=85mm \epsfbox {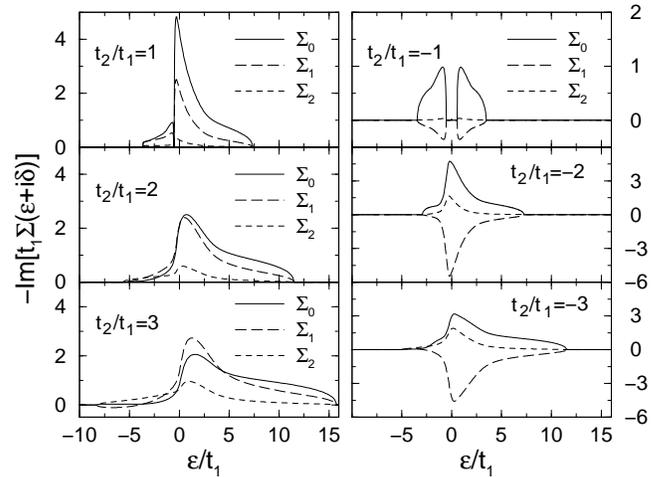}}
}\vspace{2pt}
\caption{Imaginary part of the $d$ self-energy components $\Sigma_0$,
$\Sigma_1$, and $\Sigma_2$, defined from the expansion $\Sigma_{\vec{k}} =
\Sigma_0 + \Sigma_1\epsilon_{\vec{k}} + \Sigma_2\epsilon^2_{\vec{k}}$
[see Eq.~(\protect\ref{sigma_k})], and related to the local,
nearest-neighbor, and next-nearest-neighbor self-energies through
$\Sigma_{ii} = \Sigma_0 + \Sigma_2$, $\Sigma_{<i,j>} = \Sigma_1/\sqrt{2D}$,
and $\Sigma_{<\!\!<\!i,j\!>\!\!>} = N_{<\!\!<\!i,j\!>\!\!>} \Sigma_2/2D$
[see Eqs.~(\protect\ref{self_energies})]. Here $n_f + n_d = 1$,
$U/t_1 = 4$, $T/t_1 = 1$, and $D = 2$. In the physically relevant
regime, $|t_2/t_1| \protect\agt 1$, the imaginary parts of $\Sigma_1$
and $\Sigma_0$ are comparable in size,
indicating that important nonlocal dynamical contributions are omitted
in the DMFT upon neglecting $t_2$.}
\end{figure}

The effect of CH is best seen, though, in the nonlocal $d$ self-energy
components, which are absent in the DMFT for $t_2 = 0$. Figure~4
depicts the imaginary parts of $\Sigma_0$, $\Sigma_1$, and $\Sigma_2$,
defined from the expansion $\Sigma_{\vec{k}} = \Sigma_0 +
\Sigma_1\epsilon_{\vec{k}} + \Sigma_2\epsilon^2_{\vec{k}}$. From
Eqs.~(\ref{self_energies})--(\ref{sigma_k}),
$\Sigma_1 = \sqrt{2D}\Sigma_{<i,j>}$ and
$\Sigma_2 = 2D \Sigma_{<\!\!<\!i,j\!>\!\!>}/N_{<\!\!<\!i,j\!>\!\!>}$
are the appropriately scaled nearest-neighbor and next-nearest-neighbor
self-energy components. Evidently, the imaginary part of $\Sigma_1$
(which has no Hartree contribution) is comparable to that of
$\Sigma_0$ in the physically relevant regime, $|t_2/t_1| \agt 1$.
Hence, important nonlocal dynamical contributions are omitted in the
DMFT upon neglecting $t_2$.

Anticipating a qualitatively similar result
for the Hubbard model, CH thus provides a natural mechanism for a strong
$k$-dependence in the single-particle self-energy, including for
$D \to \infty$. This may prove important, e.g., for the understanding of
the apparent $k$-dependence and the moderate mass enhancement close to
the metal-insulator transition in V$_{2-y}$O$_3$ and
Ca$_{1-x}$Sr$_x$VO$_3$.~\cite{V2O3-Kim,CaSrVO3-II} Study of
the Hubbard model along these lines is currently underway.


\begin{thebibliography}{9}
\bibitem{PH-comment} This stems from the fact that $T$ enters $G(z)$ in
      our formulation only through $n_f$ and $\mu$, which are fixed at
      $n_f = 1/2$ and $\mu = U/2$ for both $t_2 = 0$ and $t_2/t_1 = -1$.
\end{thebibliography}
\end{document}